# Diffraction grating revisited: a high-resolution plasmonic dispersive element

V. Mikhailov[1], J. Elliott[2], G. Wurtz[2], P. Bayvel[1], A. V. Zayats[2*]

[1] *Department of Electronic and Electrical Engineering, University College London,*

*Torrington Place, London WC1E 7JE, UK*

[2] *School of Mathematics and Physics, The Queen's University of Belfast, Belfast BT7*

*1NN, UK*

## Abstract

The spectral dispersion of light is critical in applications ranging from spectroscopy to sensing and optical communication technologies. We demonstrate that ultra-high spectral dispersion can be achieved with a finite-size surface plasmon polaritonic (SPP) crystal. The 3D to 2D reduction in light diffraction dimensions due to interaction of light with collective electron modes in a metal is shown to increase the dispersion by some two orders of magnitude, due to a two-stage process: (i) conversion of the incident light to SPP Bloch waves on a nanostructured surface and (ii) Bloch waves traversing the SPP crystal boundary. This has potential for high-resolution spectrograph applications in photonics, optical communications and lab-on-a-chip, all within a planar device which is compact and easy to fabricate.

The spectral dispersion of light, that is the dependence of generalised refractive index on frequency (1), is ubiquitous in applications ranging from optical spectral devices to sensing and novel optical communication technologies (*2-4*). Conventionally, both prisms and diffraction gratings can be used to disperse incident light, with gratings preferred because of their high dispersion and, thus, higher spectral resolution. Optical spectroscopy based on light-wave dispersion is employed for studying the optical properties and the internal structure of materials, atoms and molecules. Very recently, with the development in optics, new challenges have arisen for ultrahigh dispersion needed in optical switching and routing of information for wavelength-selective processing, to develop compact spectral devices that can be integrated into photonic circuits of the future or into lab-on-a-chip devices for in-situ spectral functionality in physical and life sciences applications. These cover spectral and temporal light analysis, demultiplexers, spectrometers, ultra-short pulse analysis and many others. With the developments in nanotechnology and progress in nano-fabrication, there is now a real possibility to construct nano-scale dispersive devices with uniquely tailored dispersion properties.

Metallic nanostructures possess many unusual photonic properties arising from the interaction of light with collective electron modes in a metal (*5, 6*). In the case of periodically nanostructured metallic surfaces or films, this interaction converts the "three-dimensional" incident light into "two-dimensional" surface plasmon polariton (SPP) waves propagating on a metal film surface at given frequencies, in particular directions (*6, 7*). This relies on the diffraction of light by a two-dimensional diffraction grating formed by the nanostructure at which some diffraction orders are coupled to SPP Bloch modes of the bi-grating, acting as a surface-polaritonic crystal (*6*).

Here we propose a new principle to realise a dispersive element, and report the experimental results on the nanofabricated device which uses the properties of surface plasmon polaritons on a metal film to provide angular spectral dispersion one or more orders of magnitude higher than in the state-of-the-art planar integrated diffraction gratings and photonic crystal superprism. We demonstrate this principle in a finite-size SPP crystal that can convert out-of-plane (3D) multiwavelength input into in-plane (2D) SPP output, with different light frequencies propagating in different directions. Moreover, this dispersion can be controlled optically or electrically, thus providing additional functionality in all-optical integration and signal processing. An additional advantage of the SPP approach is that no complex waveguide fabrication is required – two-dimensional guiding is inherent in the SPP propagation – significantly simplifying device fabrication over a wide wavelength range, compared to planar waveguide device technology.

The performance and spectral response of a diffraction grating depend on its period and the shape and size of the grooves determining angular dispersion and diffraction efficiency. To understand diffraction, light propagation and interaction with a grating generally should be considered in three-dimensional space (*1*). Although free-space plane and concave gratings can achieve very high resolution – their disadvantages are in that they are bulky, require discrete optical components, complex aberration control limited to a relatively narrow wavelength range, thermal stabilisation and additional components for coupling to optical fibres or waveguides, see for example (*4*). Their planar equivalents are limited to a relatively small size because of the difficulties to control the quality over a very large substrate size, suffer from higher scattering and crosstalk and lower resolution, as well as a complex waveguide fabrication. They also operate over a narrow wavelength range. Very recently two-dimensional (2D) spectral dispersion effects have been studied in two-dimensional photonic crystal slabs (*8-11*). This, so-called, superprism effect relies on the ability to direct the light transmitted through a boundary of 2D photonic crystal in different directions (i.e. with large angular separation), as a function of only very small changes in the light frequency. This effect arises from the very strong dispersion $d\vec{k}/d\omega$ of photonic eigen modes (where $\vec{k}$ is the photon wavevector and ω is the light frequency) near the edges of the photonic band-gap. Thus, the dispersion can be synthesised for a given wavelength range and required propagation direction $\vec{k}$. The photonic crystal superprism effect provides high spectral dispersion (up to 50° deviation of the light path for the incident wavelength change of 2%) but requires cumbersome arrangements for light coupling in/out of the 2D photonic crystal slab (*9,10*).

Our approach is based on the properties of surface plasmon polaritons – intrinsically 2D surface electromagnetic waves – on a periodically nanostructured metal surface forming a surface polaritonic crystal (*6*). If this structure is conventionally illuminated with a light beam, SPP eigen modes (Bloch waves), allowed to propagate on a periodic surface, will be excited at certain frequencies in specific directions at which the diffraction provides the wave vector conservation:

$$\vec{k}_{SP} = \frac{\omega}{c} n_s \sin\theta \vec{u}_{xy} \delta_p \pm p \frac{2\pi}{D} \vec{u}_x \pm q \frac{2\pi}{D} \vec{u}_y \ . \qquad (1)$$

Here, $k_{SP}$ is the SPP Bloch wave vector on the periodic structure, ω is frequency of the illuminating light, c is the speed of light, $n_s$ is the refractive index of the medium through which the film is illuminated, θ is the angle of incidence, $\vec{u}_{xy}$ is the unit vector in plane of the film in the direction of the incident light wave vector projection, $\delta_p = 1$ or 0 for p- or s-polarized incident light (relative to the sample surface), respectively; $\vec{u}_x$ and $\vec{u}_y$ are the unit reciprocal lattice vectors of the periodic structure, D is the periodicity (same in both x and y directions), and p and q are integer numbers corresponding to different directions in the SPP Brillouin zone and determining direction of SPP propagation on a surface. Thus, this scheme effectively reduces the dimension of the 3D light diffraction to 2D diffraction of surface polaritons propagating on the nanostructured surface.  Near the edges of the band-gap of SPP crystal (where $d\omega/dk_{SP} = 0$) (Fig. 1 b), the SPP wave vector can undergo extremely large variations with small changes in frequency of illuminating light.  Thus the respective SPP Bloch waves will be excited and after crossing the boundary, SPPs will emerge from the SPP crystal onto the unpatterned part of the surface in significantly different directions (Fig. 1 d). Thus, SPP crystal can convert out-of-plane multiwavelength input into in-plane SPP output with different light frequencies propagating in different directions on the smooth surface. The size and shape of nanostructure and their periods can then be optimised to achieve the required SPP dispersion and even design flat SPP bands, are ideal for high dispersion applications (*12,13*).  In addition to the wavelength-dependent effects, polarization sensitivity of the SPP coupling is also significant due to the longitudinal nature of the SPP field: the excitation field should have an electric field perpendicular to the surface or parallel to SPP propagation direction (*6,7*).

The surface polaritonic crystals used in the present work were fabricated in thin (50 nm) Au films deposited onto a glass substrates (Fig. 1a). Magnetron sputtering was used to produce high quality (small roughness is a key for minimizing background scattering of SPPs) thin metal films, in which square arrays of circular holes were fabricated with focused ion beam milling. The periodicity of the array (1500 nm) and the hole diameter (300 nm) were chosen to achieve high density of SPP Bloch mode states and flattening of SPP bands in the spectral range corresponding to optical communication wavelengths around 1.5 μm.  Near the finite size SPP crystal (50x50 μm$^2$) a set of defects was fabricated, designed to scatter SPPs into light, which is then detected in the experiments.  The defects are located at the distance of about 40 μm from the SPP crystal edge. The distance between the defect and the edge of SPP crystal allows to control linear (spectral) dispersion for the same angular dispersion provided by SPP crystal. The SPP propagation length on the Au-air surface in this wavelength range is around 200 μm and sufficient to carry energy to the defects where it is scattered in light. The defect size and shape have been chosen to provide efficient coupling of the SPP waves into light beams. For dispersion measurements, the infra-red (IR) light from a fibre-pigtailed tuneable laser was directly coupled to the SPP crystal using cleaved single mode fibre, with polarisation controlled and monitored using an all-fibre polarisation controller and polarimeter (Fig. 1c). In this way Gaussian beam illumination of the SPP crystal was obtained. As described above, the input light excites SPPs modes which are then scattered by nearby surface defect, with the resultant scattered light imaged with high-sensitivity IR camera. The recorded images at different wavelengths of the illuminating light were used to visualise the dispersive properties of the structures. The angle of incidence of the illuminating light was varied to maximise the wavelength dispersion.

The images of the resultant light distributions reveal strong frequency dependence of the intensity of SPP scattered by the defects around the SPP crystal (Fig. 2). The intensity of SPPs reaching the defects situated around the semicircle varies with the light frequency and polarization: as the frequency of the input light changes, the defects situated at different angles with respect to the structure are illuminated by the SPP beams differently, since the light of different wavelengths is coupled to SPP waves propagating in different directions on a smooth metal surface. The intensity distributions measured with unpolarized white-light illumination shows uniform distribution of scattered light around the defects apart from the central defect (D2 in Fig. 2) where the intensity is significantly lower as this defect is the closest to the corner of the SPP crystal. Lower intensity is also observed on this defect in spectral measurements (Figs. 2 and 3). The wavelength dependence of the light scattered by defects D1 and D2 (marked in Fig. 2) are shown in Fig. 3a and reveal strong changes of the scattering intensity with small changes of the frequency of light. Similarly, the cross-sections taken along the set of defects for different frequencies confirm the intensity changes on various defects with the light intensity for very small frequency variations. This shows the extremely high dispersion of the surface polaritonic crystal which is estimated from the measured images to be about $(20-30)^{o}$/nm in this spectral range. This value is an order of magnitude higher the values currently obtained with photonic superprisms $(1-2)^{o}$/nm (*9,10*) and two orders of magnitude higher than observed with one-dimensional photonic crystals and conventional wavelength division (de)multiplexing gratings (4,*14*).

Such strong dispersion in our experiments is achieved due to the two-stage diffraction process determining the dispersion of SPP crystal. Firstly, the parameters of illumination in 3D geometry (out of the SPP crystal plane) determine the direction of propagation and spatial field distribution of the SPP Bloch wave in the 2D SPP crystal. This provides the angular dispersion which can be estimated from the conditions of the excitation of the SPP waves on a smooth surface by a periodic grating to be about $0.1°/nm$ in the operating spectral range. In the case of the SPP Bloch modes' excitation in the SPP crystal this value can only be higher, as discussed above, due to the structure of allowed bands of the crystal. (Please note that this value is similar to those obtained with conventional diffraction gratings, reflecting the fact that this is a conventional diffraction process). The second stage that enhances the dispersion is the diffraction of the SPP Bloch mode on the periodic structure formed by the termination of the SPP crystal. This two-dimensional process can be considered as that of refraction on the SPP crystal boundary: the wave vector component parallel to the boundary must be conserved for the Bloch wave inside the crystal and the wave outside it (*11,15*). This is an analogue of 2D photonic crystal super-prism with both positive and negative refraction allowed, depending on the wavelength of light (Fig. 1b). The reported values for the measured superprism dispersion in 2D photonic crystals, are consistent with our measured values for the total angular dispersion of the two-stage process studied here.

In conclusion, we have proposed and measured very high spectral dispersion in finite-size SPP crystals that can convert out-of-plane multiwavelength input into in-plane SPP output with different light frequencies propagating in significantly different directions on the surface. This high dispersion occurs due to two-stage process by converting the incident light into SPP Bloch waves on a surface and then Bloch waves traversing SPP crystal boundary. The latter effect is similar to photonic superprism effect but much easier to realize and use since it does not require coupling into 2D waveguiding modes; this occurs automatically when "3D" light interacts with SPP crystal, and SPP are guided on a surface by their nature, again not requiring additional waveguides. Indeed, the combination of these effects can only be realised in an SPP crystal using surface waves (which can not be supported in a photonic crystal). This kind of SPP-induced spectral dispersion has previously never been measured, and the achieved values are very high.

These very high spectral dispersion and ease of use make the SPP crystal-based dispersive elements very promising for nano-photonic applications where wavelength splitting and separation is needed—it can be obtained in a compact integrated device which can be incorporated into lab-on-a- chip or other photonic integrated devices such as demultiplexing devices, routers, switches for optical communications. However, further device optimization is still needed since the Bloch modes used to achieve the dispersion are not plane waves and have complex, direction-dependent field distributions. This means that high-contrast-ratio spectral selection not straightforward. Also, the boundary of SPP crystal plays an important role since the deviations from perfect boundary can lead to additional splitting of the exiting beam into several beam on the smooth surface (*15*). However, we believe that the very high wavelength dispersion, which is 2-orders of magnitude higher than was previously reported, outweighs these challenges.

**Acknowledgements.**

The financial support from EPSRC is gratefully acknowledged. The authors thank to W. Dickson and R. Pollard for help with the sample fabrication and V. Tsatourian for help with computer programming.

**References.**

*Corresponding author email: a.zayats@qub.ac.uk.

**Figure captions**

Fig. 1. (a) Electron microscope image of the metallic structure used in the experiments, the insets show zooms into SPP crystal and the defects used for SPP visualisation. (b) Schematics of the band structure of the SPP crystal in the vicinity of Γ-point of the Brillouin zone, the red circle indicates the states accessible with normally incident Gaussian laser beam. (c) Experimental set-up: TLS – tuneable laser source, WLS – white light source (un-polarised), PM – power meter, OSA – optical spectrum analyser, PAN – polarisation analyser, ORC – infrared camera. (d) Schematic of the principle of wavelength dependent diffraction and the diffraction dimension reduction during light interaction with SPP crystal.

Fig. 2. Series of images of the intensity distributions over the metal surface for the different wavelengths and polarizations of the incident light as indicated on the images.

Fig. 3. (a) The wavelength dependencies of the intensity averaged over the defects D1 and D2 marked on the images in Fig. 2. (d) Cross-sections along the four scatterers as indicated in the images in Fig. 2 for different wavelengths of the incident light. Please note that due to nonlinearity of the imaging camera response, the contrast may be suppressed on the presented plots.

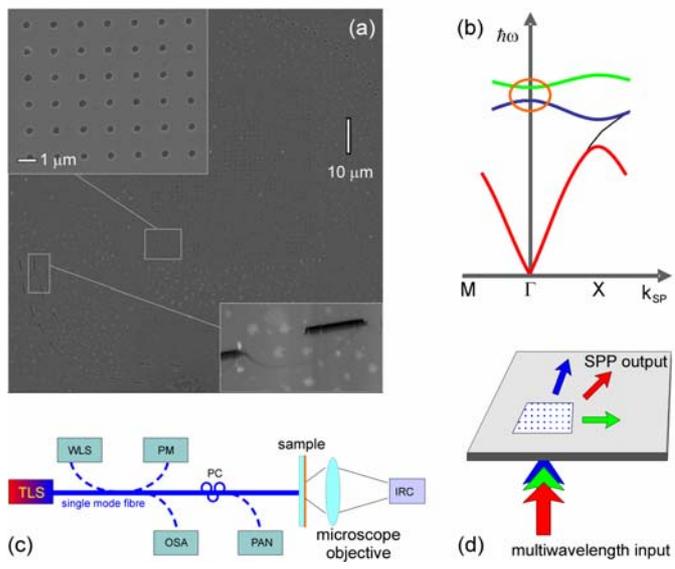

Figure 1.

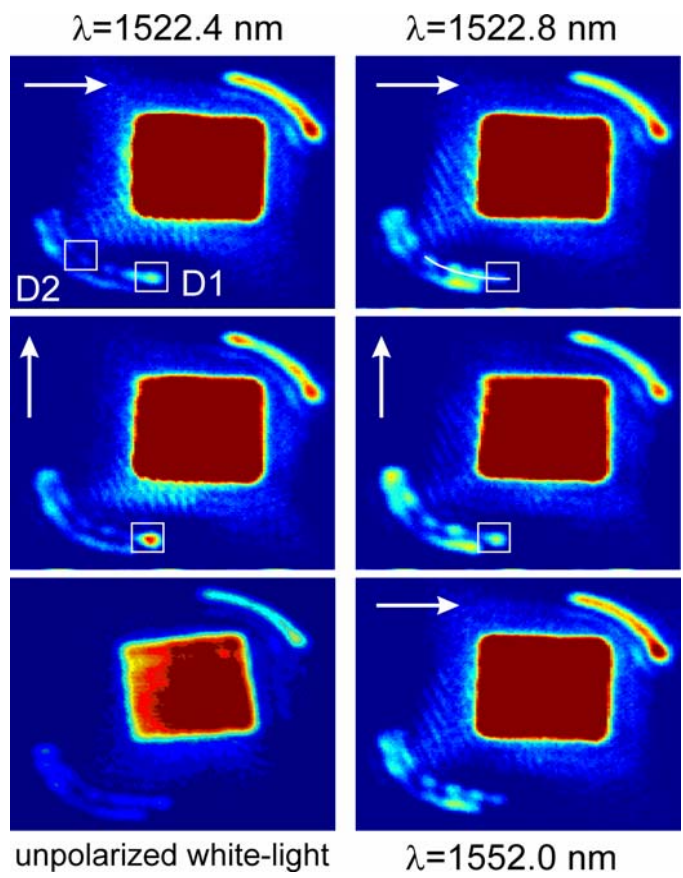

λ=1522.4 nm λ=1522.8 nm

unpolarized white-light λ=1552.0 nm

Figure 2.

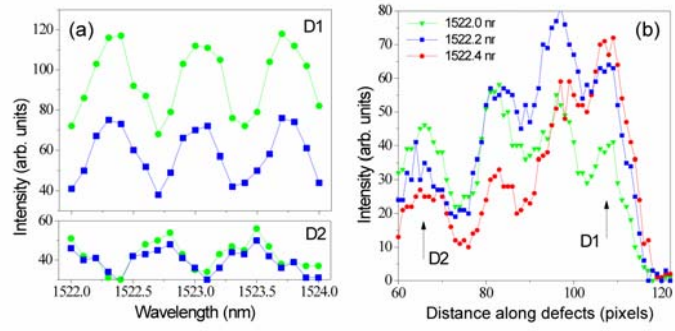

Figure 3.